\documentclass[acmlarge]{acmart}
\usepackage{adjustbox}
\usepackage{amsmath}
\usepackage{amsthm}
\usepackage{amsfonts}
\usepackage{mathrsfs}
\usepackage{mathtools}
\usepackage{bbold}
\usepackage{listings}
\usepackage{xcolor}
\usepackage{tabularx}
\usepackage{nameref}
\usepackage{graphicx}
\usepackage{xspace}
\usepackage{microtype}
\usepackage{url}
\usepackage{booktabs}
\usepackage[inline]{enumitem}
\usepackage{subcaption}
\usepackage{marvosym}
\usepackage{mathpartir}
\usepackage{stmaryrd}
\usepackage{accents}
\usepackage{bbm}
\usepackage[shortcuts]{extdash}
\usepackage{empheq}
\usepackage[T2A, T1]{fontenc}
\usepackage[utf8]{inputenc}
\usepackage[russian, english]{babel}
\usepackage{array}
\usepackage{todonotes}
\usepackage{halloweenmath}
\usepackage{cleveref}
\usepackage{textcomp}

\usepackage{etoolbox}
\usepackage{tikz}
\usetikzlibrary{calc, decorations.pathmorphing, shapes}

\setcopyright{none}
\acmPrice{}

\mathchardef\mhyphen="2D

\newcommand\para[1]{\smallskip\noindent\textbf{{#1.}\xspace}}

\makeatletter
\newcommand\tuple[1]{%
  ({#1}\@tuplehelpcheck
}
\newcommand\@tuplehelpcheck{%
  \@ifnextchar\bgroup{\@tuplehelpnext}{)}
}
\newcommand\@tuplehelpnext[1]{%
  ,\, {#1} \@ifnextchar\bgroup{\@tuplehelpnext}{)}
}
\makeatother

\makeatletter
\newcommand\pluralref[2]{%
  \ref{#1}\@pluralrefchecktwolast{#2}
}
\newcommand\@pluralrefchecktwolast[1]{%
  \@ifnextchar\bgroup{\@pluralrefcheckmore{#1}}{ and \ref{#1}}
}
\newcommand\@pluralrefcheckmore[1]{%
  \@ifnextchar\bgroup{, \ref{#1}\@pluralrefcheckmore}{, and \ref{#1}}
}
\makeatother

\makeatletter
\newlist{questions@}{enumerate}{1}

\newenvironment{questions}[1][Q]
  {\setlist[questions@]{label=#1\arabic*:, ref=#1\arabic*, font=\bfseries}
  \begin{questions@} 
  }
  {\end{questions@}
  }
\makeatother

\newcommand\sectionref[1]{\hyperref[#1]{Section~\ref*{#1}}}

\newcommand\secref[1]{\hyperref[#1]{\S\ref*{#1}}}
\newcommand\figureref[1]{\hyperref[#1]{Figure~\ref*{#1}}}

\newcommand\lemmaref[1]{\hyperref[#1]{Lemma~\ref*{#1}}}

\newcommand\theoremref[1]{\hyperref[#1]{Theorem~\ref*{#1}}}

\newcommand\appendixref[1]{\hyperref[#1]{Appendix~\ref*{#1}}}
\newcommand\lineref[1]{\hyperref[#1]{Line~\ref*{#1}}}

\newcommand\definitionref[1]{\hyperref[#1]{Definition~\ref*{#1}}}

\newcommand\tableref[1]{\hyperref[#1]{Table~\ref*{#1}}}

\newcommand\listingref[1]{\hyperref[#1]{Listing~\ref*{#1}}}

\theoremstyle{acmtheorem}
\newtheorem{theorem}{Theorem}

\theoremstyle{acmdefinition}
\newtheorem{definition}{Definition}

\makeatletter

\newcommand\makerulename[2]{%
  \expandafter\newcommand\csname def#1\endcsname{%
    \@ifundefined{r@internal@rule:explanation:#1}%
      {\Hy@raisedlink{\label{internal@rule:definition:#1}\hypertarget{rule:def:#1}{}}{#2}\xspace}%
      {\Hy@raisedlink{\label{internal@rule:definition:#1}\hypertarget{rule:def:#1}{}}\hyperlink{rule:explanation:#1}{#2}\xspace}%
  }%
  \expandafter\newcommand\csname explain#1\endcsname{%
    \Hy@raisedlink{\label{internal@rule:explanation:#1}\hypertarget{rule:explanation:#1}{}}{\hyperlink{rule:def:#1}{\LabTirName{#2}}}\xspace%
  }%
  \expandafter\newcommand\csname #1\endcsname{%
    \@ifundefined{r@internal@rule:definition:#1}%
      {\LabTirName{#2}\xspace}%
      {\hyperlink{rule:def:#1}{\LabTirName{#2}}\xspace}%
  }%
  \expandafter\newcommand\csname nolink#1\endcsname{%
    \LabTirName{#2}\xspace%
  }%
}

\makeatother

\def\C++{
  C\kern-.1667em\raise.30ex\hbox{\smaller{++}}%
  \spacefactor1000
}

\newcommand\metafun[1]{\ensuremath{\mathsf{{#1}}}}

\newcommand\toolname{\textbf{{RobustIsoCrypt}}\xspace}

\newcommand\lib{\mathtt{lib}}

\newcommand\atkclass{\mathsf{A}}

\newcommand\compileA{\mathbb{C}_{\atkclass}}

\newcommand\readonly[3]{\tuple{#1}{#2} \vdash \mathsf{read \mhyphen only}\ {#3}}

\newcommand\initialconfig[2]{\langle {#1} \mid {#2} \rangle}

\makerulename{RedCall}{red\=/call}
\makerulename{RedReturn}{red\=/ret}
\makerulename{RedBeta}{red\=/$\beta$}

\makerulename{BetaSubst}{$\beta$\=/subst}
\makerulename{BetaOp}{$\beta$\=/op}
\makerulename{BetaDeref}{$\beta$\=/deref}
\makerulename{BetaDerefOOB}{$\beta$\=/deref\=/oob}
\makerulename{BetaWrite}{$\beta$\=/write}
\makerulename{BetaWriteOOB}{$\beta$\=/write\=/oob}
\makerulename{BetaNew}{$\beta$\=/new}
\makerulename{BetaGetBlock}{$\beta$\=/get\=/block}
\makerulename{BetaGetOffset}{$\beta$\=/get\=/offset}
\makerulename{BetaIfFalse}{$\beta$\=/if\=/false}
\makerulename{BetaIfTrue}{$\beta$\=/if\=/true}
\makerulename{BetaProtect}{$\beta$\=/protect}
\makerulename{BetaUnprotect}{$\beta$\=/unprotect}
\makerulename{BetaSeq}{$\beta$\=/seq}

\makerulename{SpecProtect}{spec\=/protect}
\makerulename{SpecNonspec}{spec\=/nonspec}
\makerulename{SpecSpec}{spec\=/spec}
\makerulename{SpecFence}{spec\=/fence}
\makerulename{SpecNoMispecVal}{spec\=/final\=/val}
\makerulename{SpecFenceVal}{spec\=/fence\=/val}

\makerulename{FenceNoMispec}{fence\=/no\=/mispec}
\makerulename{FenceRollback}{fence\=/rollback}

\makerulename{AttackerReadOnly}{app\=/read\=/only}
\makerulename{AttackerReadOnlyFinal}{app\=/read\=/only\=/end}
\makerulename{AttackerMemSafe}{app\=/mem\=/safe}
\makerulename{AttackerMemSafeFinal}{app\=/mem\=/safe\=/end}
\makerulename{TraceReadOnly}{trace\=/read\=/only}
\makerulename{TraceMemSafe}{trace\=/mem\=/safe}

\newcounter{criteria}
\crefname{criteria}{}{}

\newcommand\tocite[1]{\textcolor{red}{[??]}}
\definecolor{Ckeywordblue}{RGB}{36, 75, 133}
\definecolor{Ccommentbrown}{RGB}{142, 89, 19}
\definecolor{Csymbolorange}{RGB}{205, 92, 25}
\definecolor{Cconstantblue}{RGB}{7, 27, 203}

\lstdefinestyle{C}{%
  numbers=left,
  numberstyle=\tiny,
  basicstyle=\footnotesize\ttfamily,
  columns=fullflexible,
  keepspaces=true,
  otherkeywords={*, =},
  keywords=[1]{int, void, static, unsigned, char, const},
  keywordstyle=[1]\color{Ckeywordblue}\bfseries,
  keywords=[2]{*, =},
  keywordstyle=[2]\color{Csymbolorange}\bfseries,
  comment=[l]{//},
  commentstyle=\color{Ccommentbrown},
  xleftmargin=2.5em
}

\newcommand\Cinline[1]{\lstinline[style=C, basicstyle=\small\ttfamily]{#1}}

\newcommand\eg{e.g.,\xspace}

\begin{document}

\title[Robust Constant-Time Cryptography]{Robust Constant-Time Cryptography}

\newcommand\ucsdmark{*}
\newcommand\mpimark{\textdagger}
\newcommand\trentomark{\ddag}
\newcommand\utrechtmark{\S}

\author{Matthew Kolosick}
\affiliation{%
\institution{University of California San Diego}
\country{USA}
}

\author{Basavesh Ammanaghatta Shivakumar}
\affiliation{%
\institution{Max Planck Institute for Security and Privacy}
\country{Germany}
}

\author{Sunjay Cauligi}
\affiliation{%
\institution{Max Planck Institute for Security and Privacy}
\country{Germany}
}

\author{Marco Patrignani}
\affiliation{%
\institution{University of Trento}
\country{Italy}
}

\author{Marco Vassena}
\affiliation{%
\institution{Utrecht University}
\country{Netherlands}
}

\author{Ranjit Jhala}
\affiliation{%
\institution{University of California San Diego}
\country{USA}
}

\author{Deian Stefan}
\affiliation{%
\institution{University of California San Diego}
\country{USA}} 


\authorsaddresses{}
\renewcommand{\shortauthors}{M. Kolosick, B. Shivakumar, S. Cauligi, M. Patrignani, M. Vassena, R. Jhala, D. Stefan}

\maketitle

\section{Extended abstract}

The constant-time property is considered the security standard for
cryptographic code.
Code following the constant-time discipline is free from
secret-dependent branches and memory accesses, and thus avoids leaking
secrets through cache and timing
side-channels~\cite{cauligi_sok_2022,barthe_system-level_2019}.
Though security against side-channel attacks is an important concern
for secure cryptographic implementations~\cite{brumley_remote_2005},
the constant-time property makes a number of implicit assumptions that
are fundamentally at odds with the reality of cryptographic code.

\para{Constant-time is not robust}
The first issue with constant-time is that it is a
\emph{whole-program} property: It relies on the \emph{entirety} of the
code base being constant-time.
But, cryptographic developers do not generally write whole programs;
rather, they provide libraries and specific algorithms for other
application developers to use.
As such, developers of security libraries must maintain their security
guarantees even when their code is operating within (potentially
untrusted) application contexts.

\para{Constant-time requires memory safety}
The whole-program nature of constant-time also leads to a second
issue: constant-time \emph{requires} memory safety of \emph{all} the
running code.
\emph{Any} memory safety bugs, whether in the library or the
application, will wend their way back to side-channel leaks of secrets
if not direct disclosure.
And although cryptographic libraries should (and are) written to be
memory-safe, it's unfortunately unrealistic to expect the same from
every application that uses each library.

We provide an example from the libsodium cryptographic
library~\cite{bernstein2019libsodium}: The code below shows the
(abridged) C implementation of the Salsa20 stream cipher, a
constant-time encryption primitive.
\begin{lstlisting}[style=C, escapeinside=||]
static int stream_ref(u8 *c, u64 clen, u8 *n, u8 *k)
{
    ... u8 kcopy[32]; ...
    for (i = 0; i < 32; i++) {
        kcopy[i] = k[i];
    }
    ...
    while (clen >= 64) {
      crypto_core_salsa20(c, in, kcopy, NULL); ...
    }
    ...
    sodium_memzero(kcopy, sizeof kcopy); |\label{code:salsa-ref:memzero}|
    return 0;
}
\end{lstlisting}
\noindent
The local buffer \Cinline{kcopy} holds a copy of the key, which is
left unchanged by \Cinline{crypto_core_salsa20}.
The data held in \Cinline{kcopy} is not returned from the function,
and (since the algorithm is constant-time) it is not leaked through
any timing side-channels, so it should not matter if its contents are
wiped or not.
%
However, if an attacker is able to, \eg exploit a memory safety
vulnerability in the linked application code, they might be able to
read arbitrary (or targeted) bits from memory, allowing them to steal
the key data from the leftover \Cinline{kcopy}.
Clearly, the classical definition of constant-time security is
insufficient to capture this notion.

\para{Different attackers require different defenses}
While libraries like libsodium add memory clearing defenses like those
shown above, others choose to elide them.
We argue that neither of these choices is universally correct: For
example if libsodium is run in a safe Rust application, clearing the
intermediate memory is unnecessarily defensive; whereas if a linked
application contains memory disclosure bugs, then a library without
such mitigations will be leaving sensitive data vulnerable to an
attacker.
Ideally, software security properties for cryptographic code should
allow us to reason about which protections are needed based on the
what kinds of applications it will be linked with.

\para{Spectre complicates everything}
Finally, albeit quite unsurprisingly, \emph{speculative execution}
complicates even further any discussion about cryptographic software
security.
Just as with classical constant-time, we already have a variety of
tools and formal techniques to ensure that cryptographic code
\emph{itself} is protected from Spectre
attacks~\cite{cauligi_sok_2022}. However, these techniques usually
come with substantial performance tradeoffs, making them impractical
to use for the entirety of an application.
Due to Spectre attacks, applications that appear memory-safe can still
be tricked into revealing arbitrary memory data; once again
cryptographic libraries must not only harden their own code, but must
also defend against these \emph{speculative} vulnerabilities in the
application.
Although there has been much work developing constant-time properties
for speculative execution~\cite{cauligi_sok_2022}, speculative
constant-time is still a whole-program property; it, too, is wholly
insufficient if we want to make guarantees for cryptographic
libraries.

\subsection{Robust constant-time}

Our answer to this problem is
\emph{robust constant-time}.
Like other robustness
properties~\cite{rsc-j,rs-prots,rs-types,catalin-rs,rs-caps,rs-sandboxing},
robust constant-time ensures that a given library does not leak
secrets \emph{regardless of the linked application}.
In addition, we capture the varying assumptions about
applications---memory safety, the presence of read gadgets,
speculation, etc.---as \emph{classes} of application contexts.
This allows us to formally examine different mitigation strategies
when linking against, for instance, safe Rust applications, buggy C
applications, or even applications full of Spectre gadgets.
Our notion of robust constant-time underpins our formal security model
for developing cryptographic libraries: Not only must a library be
secure, it must remain secure even in a given attacker context.

To formally define robust constant-time, we first define
\emph{libraries}, which we parameterize with a set of \emph{API
functions} $\Gamma$ and a set of \emph{secrets} $\Delta$:
\begin{definition}[$\Gamma$-$\Delta$-libraries]
  Given an API context $\Gamma$ and secret context $\Delta$, we say $L$ is a
  $\Gamma$-$\Delta$-\emph{library} with private context $\Gamma_p$ when $L$ is a
  closing substitution for $\Gamma_p \uplus \Gamma$ with codomain $\lib$-labeled
  functions well-formed under $\Delta$.
\end{definition}

\subsection{Characterizing attackers}

Since application/attacker assumptions differ, we also parameterize
our definition of robust constant-time with a \emph{class of contexts}
to capture the variation in attacker models.
For instance, we can assume an application written in safe Rust is
memory safe, and thus we don't need to worry about memory disclosure
bugs from the application. We can thus apply different protections to
our library than if we were linking against applications written in C
or where Spectre gadgets are a concern.

We formally define an \emph{attacker} in terms of a \emph{trace safety
property} $\Gamma \vdash e$ which captures the set of operations the
attacker/application is allowed make when interacting with the
library.
The four concrete attack classes we define are:
\begin{enumerate*}[label=\arabic*)]
\item Memory-safe attackers, where the application can neither read nor write out-of-bounds;
\item read-only attackers, where the application might read out-of-bounds;
\item memory-unsafe attackers, where the application contains arbitrary memory safety vulnerabilities; and
\item speculative attackers, where the application contains Spectre-style vulnerabilities.
\end{enumerate*}
We define speculative attackers via a novel, high-level speculative
semantics, parameterized by a \emph{speculator} that controls when and
how speculation and rollback occur.
This definition of speculation allows us to capture a wide variety of
possible speculative attacks.

As an example, our formal definition of a read-only attacker is as
follows:
\begin{definition}[Read-only attacker]
$\Gamma \vdash e$ is a \emph{read-only attacker} if, for all $\Gamma$-$\Delta$
libraries $L$, initial states $S \vDash \Delta$, and $\overline{\alpha} \in
 \metafun{traces}(\initialconfig{S}{L(e)})$, we have $\readonly{\emptyset}{\emptyset}{\overline{\alpha}}$.
\end{definition}
\noindent
The notation $\readonly{\emptyset}{\emptyset}{\overline{\alpha}}$ captures that the
only ``bad'' actions $e$ performs are reading out-of-bounds;
the definition of $\mathsf{read \mhyphen only}$ partitions
the trace into alternating sequences of application and library events.
We impose restrictions on the application events under the assumption that
the library events are well-behaved, in essence treating the library as a
``context'' for executing the application.

We can now define robust constant-time in the context of read-only attackers:
\begin{definition}[Read-only robust constant-time]
  We say a $\Gamma$-$\Delta$-library $L$ is \emph{read-only robustly constant-time}
  if, for all read-only attackers $\Gamma \vdash e$ and secret states $S$
  and $S'$ such that $S \vDash \Delta$ and $S' \vDash \Delta$,
  we have
  $\operatorname{traces}(\initialconfig{S}{L(e)}) =
  \operatorname{traces}(\initialconfig{S'}{L(e)})$.
\end{definition}
\noindent
We can define the other attacker models similarly. For speculative
execution, we further parameterize by a class of speculators and we
relate the \emph{speculative} traces (those produced by the
speculative semantics).

\subsection{A robust constant-time compiler}

Robust constant-time makes implicit security guarantees concrete.
For example, with robust constant-time we can show which functions
\emph{must} clear temporary data---like in the libsodium example---%
and which functions can get away without additional mitigations.
Unfortunately, current cryptographic libraries implement these mitigations manually.
As such, they fundamentally limit themselves to a single attacker
model, and must make compromises in their level of security---always
picking the strongest possible protections would lead to significant
(and unnecessary) overhead.
For example, the libsodium developers explicitly chose not to add
speculative protections,\footnote{libsodium forgoes an appropriate
memory fence in their implementation of \Cinline{sodium_memzero}:
\url{https://github.com/jedisct1/libsodium/issues/802}.} instead
opting for best-effort protections in such cases to maintain
performance.

We instead build a compiler that is aware of robust constant-time and
is \emph{parameterized by the attacker model}, allowing us to compile
a library with different mitigations for different attacker models.
Thus the same library code can be used safely and efficiently whether
called from Rust or from C, or even when Spectre is a concern.
During compilation, our compiler performs a static taint analysis of
the library code to determine which data is secret or can reveal
secret information.
Then, depending on the attack model, the compiler moves secret data to
protected memory regions, clears accessible temporary secrets, and
sanitizes context switches between the library and application code.

Formally, we assume the cryptographic library is already classically
constant-time; this can be achieved through existing tools such as
Blade~\cite{blade}.
For each attacker class $\atkclass$, we then have a compiler
$\compileA$ such that compiling a constant-time library $L$ with
$\compileA$ guarantees that it is robustly constant-time against
$\atkclass$:
\begin{theorem}[Compiler is secure]
  Let $L$ be a $\Gamma$-$\Delta$-library such that $L$ is classically constant-time.
  Then $\compileA(L)$ is robustly constant-time w.r.t. attacker class $\atkclass$.
\end{theorem}

\section{A robust compiler}
\label{sec:robust-compiler}

\begin{figure}[H]
  \centering
  \includegraphics[width=12cm,keepaspectratio]{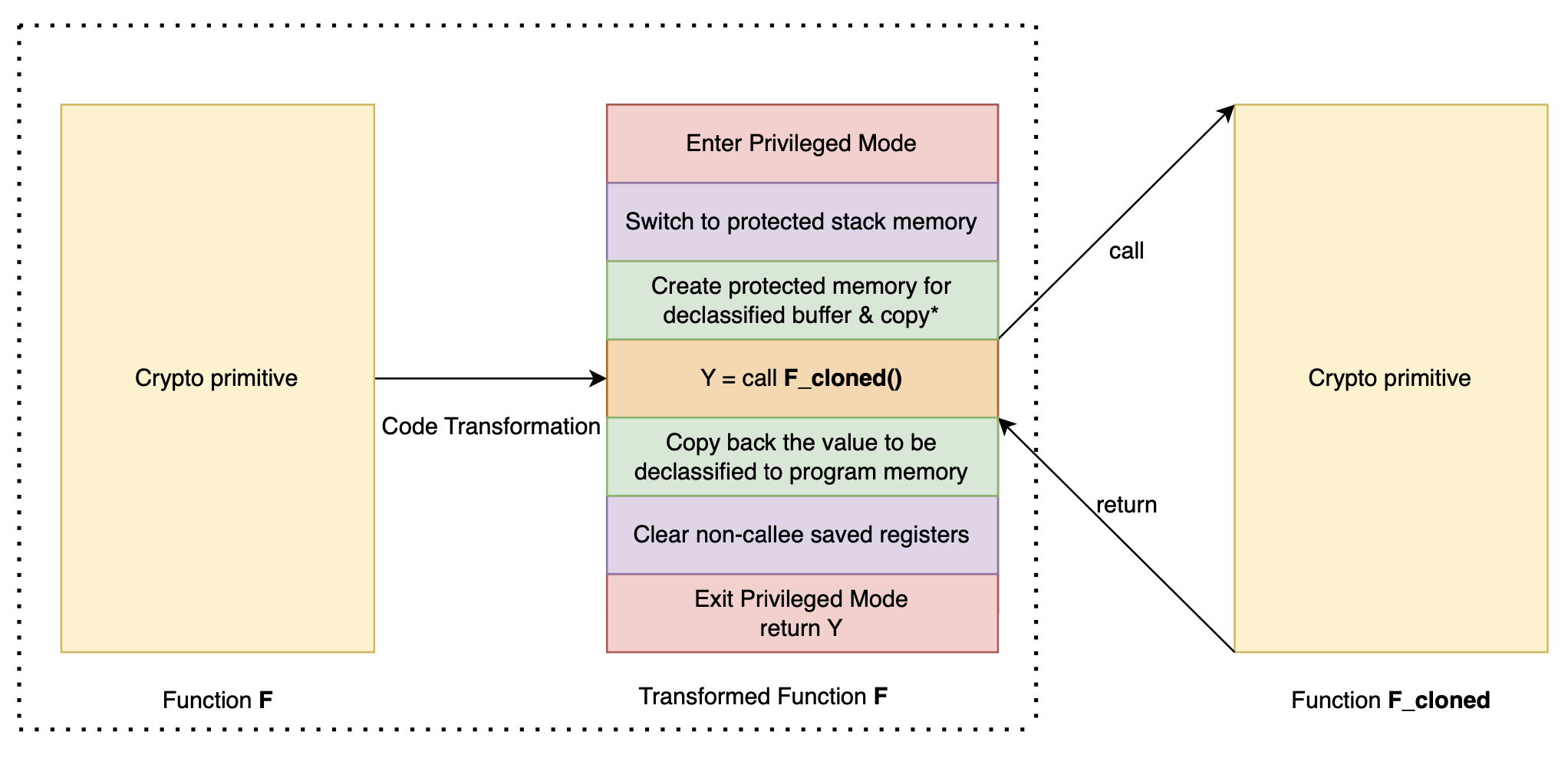}
  \caption{\toolname{} wrapping of cryptographic API functions}
  \label{fig:libfuncwrapper}
\end{figure}

We develop \toolname{}, a compiler for providing robust (speculative) constant
time protections for cryptographic libraries.
Built on top of the LLVM compiler framework~\cite{lattner2004llvm}, \toolname{} uses
Intel's Memory Protection Keys (MPK) to guarantee that secret data
(cryptographic secrets and the information derived from them) is only accessible
while executing trusted cryptographic library code.
The use of MPK ensures that, not only is this secret data inaccessible while
executing within the untrusted application, but it also guarantees that all data
within the MPK protected region is inaccessible from parallel (potentially
untrusted) threads within the same application.
We discuss the design of the robust library protections in
\sectionref{sec:robust-compiler:library}.

While the robust library protections guarantee that the cryptographic code
executes within the protected domain, cryptographic secrets generally originate
and are managed by application code.
As such we design an LLVM pass that compiles client code to ensure that
cryptographic secret buffers are allocated within protected memory.
We discuss the design of this client pass in
\secref{sec:robust-compiler:application}.

\subsection{Making libraries robust}
\label{sec:robust-compiler:library}

Cryptographic code operates directly on secret data and, to prevent timing-based
leaks, is required to be constant time.
Given these baseline security requirements we operate under the assumption that
cryptographic code is \emph{trusted}.
The task of \toolname{} then is to ensure that the secret data operated on by
cryptographic code is inaccessible from untrusted client code.
These protections are provided in three steps:
First, cryptographic developers label the external API functions, those
functions that will be directly called by the client.
Second, \toolname{} wraps these API functions to handle the memory domain switching.
Thirdly, \toolname{} replaces all dynamic memory functions (\Cinline{malloc} and
similar) with custom MPK compatible versions, ensuring that all memory allocated
within the cryptographic library is kept within the protected domain.

\figureref{fig:libfuncwrapper} shows our wrapping of cryptographic API functions.
For every exported function \Cinline{F} in the library, a corresponding clone,
\Cinline{F_clone}, is generated that contains the original implementation of \Cinline{F}.
All \emph{internal} calls to \Cinline{F} are then replaced with calls to
\Cinline{F_clone} and the new \Cinline{F} becomes the external API wrapper for
use by the client.
\Cinline{F} is responsible for switching to the protected memory region and
switching to a protected stack within that memory region.

The new \Cinline{F} takes the following domain switching steps:
\begin{enumerate*}
\item Access to the protected memory region is enabled with a \Cinline{wrpkru}
instruction.

\item We get the address of the stack within the protected memory region and we
save the current stack pointer to the top of the protected stack.

\item We copy stack arguments to the new protected stack. 

\item If the parallel protections are enabled, then we allocate a copy buffer
for external buffers (we discuss this below).

\item The stack pointer is switched to the protected stack frame.

\item We call \Cinline{F_clone}.

\item Any copied buffers are written to the original buffer.

\item We clear all scratch registers.

\item We exit disable access to protected memory and return to the client. 
\end{enumerate*}
Together these ensure that all data produced and used by the cryptographic code
is within the protected memory region and the region is inaccessible to
application code.

\begin{figure}
  \centering
  \includegraphics[width=12cm,keepaspectratio]{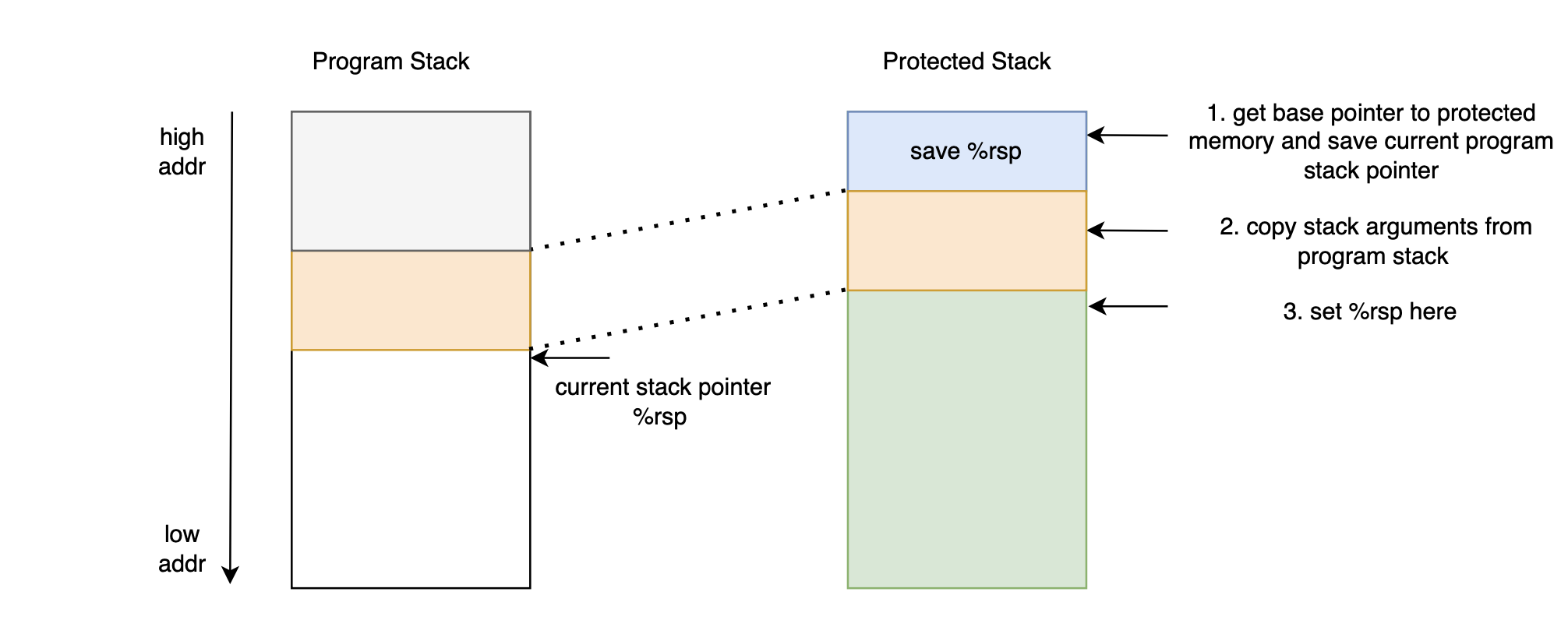}
  \caption{\label{fig:stack-stitching}%
  Switching from program stack to protected stack}
\end{figure}

\para{Parallel protections}
Rather than allocate extra memory, cryptographic algorithms will sometimes carry
out intermediate computations within output (often ciphertext) buffers.
In a single-threaded context, this is safe as there is no way for client code to
access these buffers before they contain their final, declassified (cryptographically secure)
value.
In a parallel context, work like
Spectre-Declassfied~\cite{shivakumar2023spectre} has shown that an attacker can
recover secret information if they can observe the intermediate results before
the value is ready to be declassified.
To defend against these attacks we add a parallel protection option that
allocates memory for these intermediate buffers within protected memory,
performs the intermediate work within the protected domain, and then copies the
declassified result back out to the unprotected memory.

\subsection{Client Pass}
\label{sec:robust-compiler:application}

While the cryptographic library manages all cryptographic operations,
the secret key, plaintext, and other essential input parameters
originate from the client code. Hence, it's imperative to guarantee
that crypto buffers are allocated within the protected memory domain
and accessed in privileged mode to the minimum extent possible from
the client code.

We use CryptoMPK's~\cite{jin2022annotating} mitigation strategy to
track the secret information data flow to rellocate crypto buffers to
protected memory region and perform privilege switch instrumentation.
In our methodology, the function calls to the library are considered
foreign function calls and the library's source code is not available
to the client pass. Hence, the data flow tracking becomes much simpler
as the analysis does not go into deep nested calls.

\section{Evaluation}
\label{sec:eval}

\begin{center}
  \input{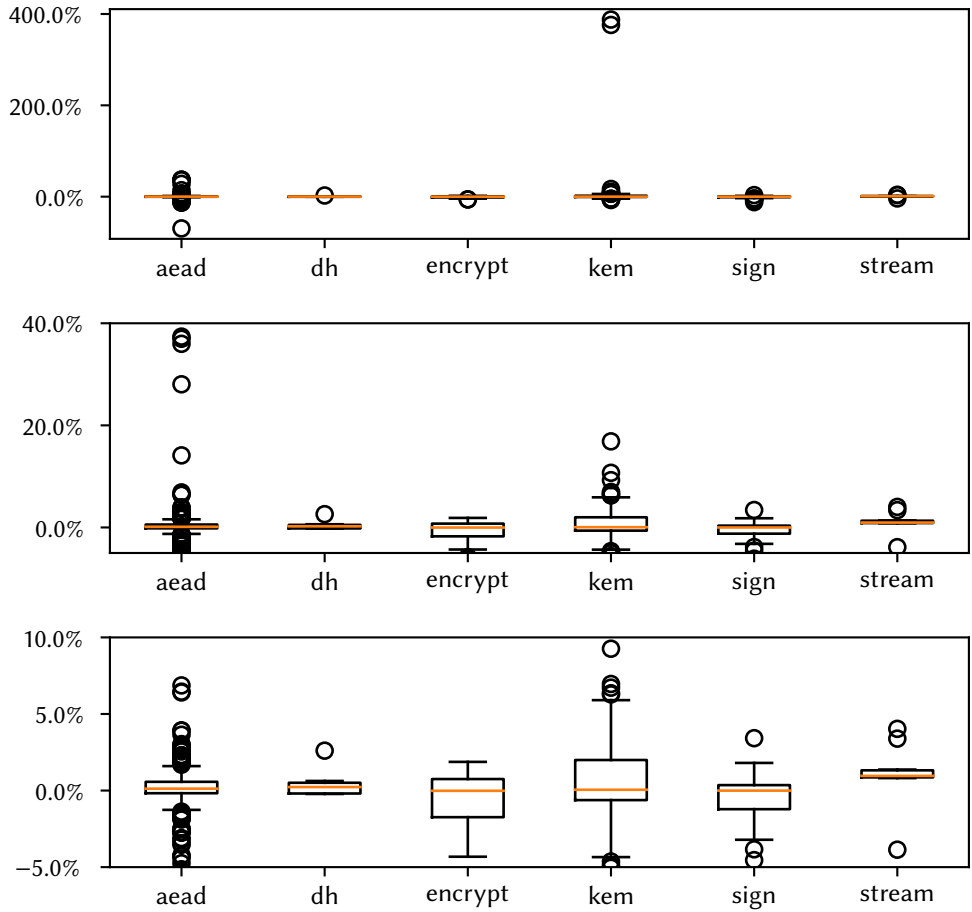}
  \captionof{figure}{Percent overheads of \toolname{} for SUPERCOP operations.}
  \label{fig:robust-overheads}
\end{center}

To evaluate the cost of guaranteeing robust constant time we ask the following questions:
\begin{questions}
\item What is the overhead of guaranteeing \emph{robust} constant against
speculative/read-only attackers?\footnote{Robust protections against speculative
attackers require the additional clearing of scratch registers, however the
overhead of this is low-enough that we do not omit the clearing in our
non-speculative protections.}

\item What is the overhead of guaranteeing \emph{robust} constant time against
in-process, parallel attackers?

\end{questions}

\para{Benchmarks}
In parallel to the narrow focus on the properties of constant and speculative
constant time, there is an unfortunate trend in \emph{evaluating} cryptographic
software security techniques on the same small set of benchmarks.
These benchmarks are typically limited to some implementation of stream ciphers
from the Salsa20/Chacha20 families, Poly1305, the SHA hash functions, AES, and
Curve25519.
The implementations further typically come from libsodium~\cite{bernstein2019libsodium},
OpenSSL~\cite{openssl}, and small verified crypto libraries such as
HACL*~\cite{zinzindohoue2017hacl} or Jasmin~\cite{almeida2017jasmin}.
While these are indeed some of the most common cryptographic algorithms and
implementations (and, in the case of the verified implementations, ones well
worth studying and promoting), they are a small glimpse into the variations of
cryptographic algorithms and implementations.
In an attempt to remedy this limitation we modify the SUPERCOP~\cite{supercop}
cryptographic benchmarking tool so that it can be used to study the costs of
applying protections to cryptographic algorithms.
The SUPERCOP benchmarks contain a wide array of different cryptographic
algorithms and implementations thereof (for instance our benchmarks are run on
over 300 AEAD algorithms).
Due to the nature of our LLVM implementations, we do limit our benchmarking suite
to the pure C/C++ implementations of the algorithms.

To answer our research questions we evaluate each intervention on the AEAD,
Diffie Hellman, public key encryption, key encapsulation mechanisms, public key
signatures, and stream cipher benchmark suites within SUPERCOP.

\para{Machine and software setup}
We run all benchmarks on a 13th Gen Intel\textregistered{} Core\texttrademark{}
i9-13900KS, 125GB RAM, running Linux kernel version 6.3.0.
We run SUPERCOP with its configuration to collect data only on cores with the
same frequency and all of our data is collected from CPUs running at 5.6 GHz.
\toolname{} adds new passes to LLVM 16.0.2 and is split into two passes: the
library pass which adds the bulk of the protections and the application pass
which allocates a stack in protected memory at the start of \Cinline{main}.
SUPERCOP defines API functions for each operation and these are annotated as the
external entry points to the cryptographic libraries.
The parallel protection benchmarks label the ciphertext arguments on the
Diffie-Hellman and stream APIs as \emph{declassify}.
We use a modified version of jemalloc~\cite{evans2006scalable} patched to provide MPK
allocation functions.
Our baseline replaces the libc malloc implementation with an unpatched version
of jemalloc.

\para{Summary of results}
We find that robust constant time protections can be generally be guaranteed
with minimal overhead (less than 5\% in almost all cases), though there do exist
outliers with up to 20\% for key encapsulation mechanisms and 40\% overhead for
AEAD algorithms as well as one algorithm in the key encapsulation mechanism data
set with around 400\% overhead.
At small data sizes highly optimized stream ciphers also pay a large overhead
percent, however on these workloads the baseline workload takes the order of a
few hundred cycles.
There is a higher, but still minimal overhead for protecting against parallel
attackers.

\begin{center}
  \captionof{table}{%
    Overheads for robust protections compared to unprotected baseline.
  }
  \begin{tabular}{l|cccccc}
    \toprule
    \textbf{Benchmark}
    & \textbf{aead}
    & \textbf{dh}
    & \textbf{encrypt}
    & \textbf{kem}
    & \textbf{sign}
    & \textbf{stream}
    \\
    \toprule
    N
    & 385
    & 9
    & 15
    & 84
    & 38
    & 11
    \\
    Data size (bytes)
    & 2048002048
    & fixed
    & 96397
    & fixed
    & 96397
    & 4096
    \\
    Median overhead
    & 0.12\%
    & 0.22\%
    & 0.07\%
    & 0.07\%
    & 0.19\%
    & 1.11\%
    \\
    IQR
    & 0.67\%
    & 0.49\%
    & 2.23\%
    & 2.64\%
    & 1.07\%
    & 0.39\%
    \\
    Baseline median cycles
    & 3.09e+05
    & 1.57e+05
    & 1.42e+08
    & 7.80e+05
    & 3.00e+08
    & 1.54e+04
    \\
    \bottomrule
  \end{tabular}
\end{center}

\begin{center}
  \captionof{table}{%
    Overhead for stream ciphers with varying plaintext sizes.
  }
  \begin{tabular}{l|cccc}
    \toprule
    \textbf{Data size}
    & \textbf{Median overhead}
    & \textbf{IQR}
    & \textbf{Baseline median cycles}
    & \textbf{Median overhead cycles}
    \\
    \toprule
    1 & 30.72\% & 31.87\% & 4.08e+02 & 143 \\
    128 & 16.4\% & 10.51\% & 8.02e+02 & 116 \\
    256 & 10.41\% & 7.15\% & 1.40e+03 & 129 \\
    512 & 5.87\% & 3.13\% & 2.12e+03 & 119 \\
    1024 & 3.59\% & 2.35\% & 4.08e+03 & 131 \\
    2048 & 1.93\% & 1.42\% & 7.72e+03 & 121 \\
    4096 & 1.11\% & 0.39\% & 1.54e+04 & 139 \\
    \bottomrule
  \end{tabular}
\end{center}

\begin{center}
  \captionof{table}{%
    The cost of parallel protection: dh
  }
  \begin{tabular}{cccc}
    \toprule
    \textbf{median}
    & \textbf{declassify median}
    & \textbf{IQR}
    & \textbf{declassify IQR}
    \\
    \toprule
    0.22\%
    & 0.41\%
    & 0.49\%
    & 0.66\%
    \\
    \bottomrule
  \end{tabular}
\end{center}

\begin{center}
  \captionof{table}{%
    The cost of parallel protection: stream
  }
  \begin{tabular}{l|cccc}
    \toprule
    \textbf{Data size}
    & \textbf{median}
    & \textbf{declassify median}
    & \textbf{IQR}
    & \textbf{declassify IQR}
    \\
    \toprule
    1 & 30.72\% & 41.18\% & 31.87\% & 39.12\% \\
    128 & 16.4\% & 19.88\% & 10.51\% & 15.95\% \\
    256 & 10.41\% & 11.93\% & 7.15\% & 10.12\% \\
    512 & 5.87\% & 8.35\% & 3.13\% & 4.64\% \\
    1024 & 3.59\% & 5.60\% & 2.35\% & 3.69\% \\
    2048 & 1.93\% & 2.95\% & 1.42\% & 1.78\% \\
    4096 & 1.11\% & 1.62\% & 0.39\% & 1.48\% \\
    \bottomrule
  \end{tabular}
\end{center}

\bibliography{local.bib}

\end{document}